# How to define and test an Indirect Moderation model: the missing link in regression-based path models.

Geert H. van Kollenburg [a*] & Marcel A. Croon [b]

[a] Interconnected Resource-aware Intelligent Systems, Department of Mathematics and Computer Science, Eindhoven University of Technology, Eindhoven, The Netherlands.

[b] Department of Methodology and Statistics, Tilburg School for the Social and Behavioural Sciences, Tilburg University, Tilburg, The Netherlands

* Corresponding author information: Geert van Kollenburg, Den Dolech 2, 5612 AZ Eindhoven, The Netherlands (e-mail: g.h.v.kollenburg@tue.nl).

## Abstract

Two of the most important extensions of the basic regression model are moderated effects (due to interactions) and mediated effects (i.e. indirect effects). Combinations of these effects may also be present. In this work, an important, yet missing combination is presented that can determine whether a moderating effect itself is mediated by another variable. This 'indirect moderation' model can be assessed by a four-step decision tree which guides the user through the necessary regression analyses to infer or refute indirect moderation. A simulation experiment shows how the method works under some basic scenarios.

Keywords: Regression analysis, Mediation, Moderation, Structural Equation Modelling

## 1. Introduction

Regression analysis is arguably one of the most common statistical methods in social scientific research and predictive. Many applications extend the basic regression model with moderation (i.e., interaction) effects by including product terms of predictor variables into the model (James & Brett, 1984). By combining regression models into a (simple) path model, mediation (i.e., indirect) effects can be tested (Baron & Kenny, 1986; Hayes, 2018), as an alternative to using structural equation modelling (Keith, 2019).

While assessment of mediation and moderation is well established (i.e., Hayes, 2009; Mallinckrodt *et al.*, 2006), methods for testing integrations of mediation and moderation have been under debate for some time (Edwards & Lambert, 2007; Morgan-Lopez & Mackinnon, 2006; Muller et al., 2005; Ye & Wen, 2013). Integrated models of mediation and moderation have to date only focused only on extending the mediation model. While these moderated mediation models are interesting for specific applications, a void has remained in the literature when it came to incorporating mediation into the moderation model. In currently available models, indirect effects can be moderated, but no models exist in which the moderation is indirect.

This tutorial presents the indirect moderation model, which can be estimated using a multiple regression analysis. The next section will explain the basic methodology of mediation and moderation analysis. After that, integration of mediation and moderation are discussed. Special attention is given to the model named 'mediated moderation'. After that, the indirect moderation model is presented and will be compared to the mediated moderation model in an exploratory fashion to illustrate the uniqueness of both models. An easy-to-use 4-step decision tree is presented which guides researchers towards assessment of indirect moderation. A discussion of the results and implications for moderation analysis will be presented at the end.

## 2. **Theory and methods**

### *2.1 Path modelling using linear regression*

Linear regression is the most common version of regression analysis (and the only one this report will focus on). In its simplest form, it can be used to estimate the relationship between a dependent variable (also called target, or outcome) and a predictor variable (also called independent variable, or feature; the term covariates is often used to indicate control variables). A regression model is a supervised prediction model, which estimates the relationship between some predictor variable(s) X and a target variable Y. The resulting regression model can be used to predict unknown Y values from observed X observations.

Figure 1 shows a scatter plot of 200 pairs $\{x, y\}$. The red diagonal line is the best fitting line (based on least squares estimation). The basic regression model for predicting $Y$ from $X$ is

$$Y = \beta_{10} + \beta_{1X}X + \varepsilon_1 \qquad (1)$$

Which can be conceptually represented as a path model, shown in Figure 2.

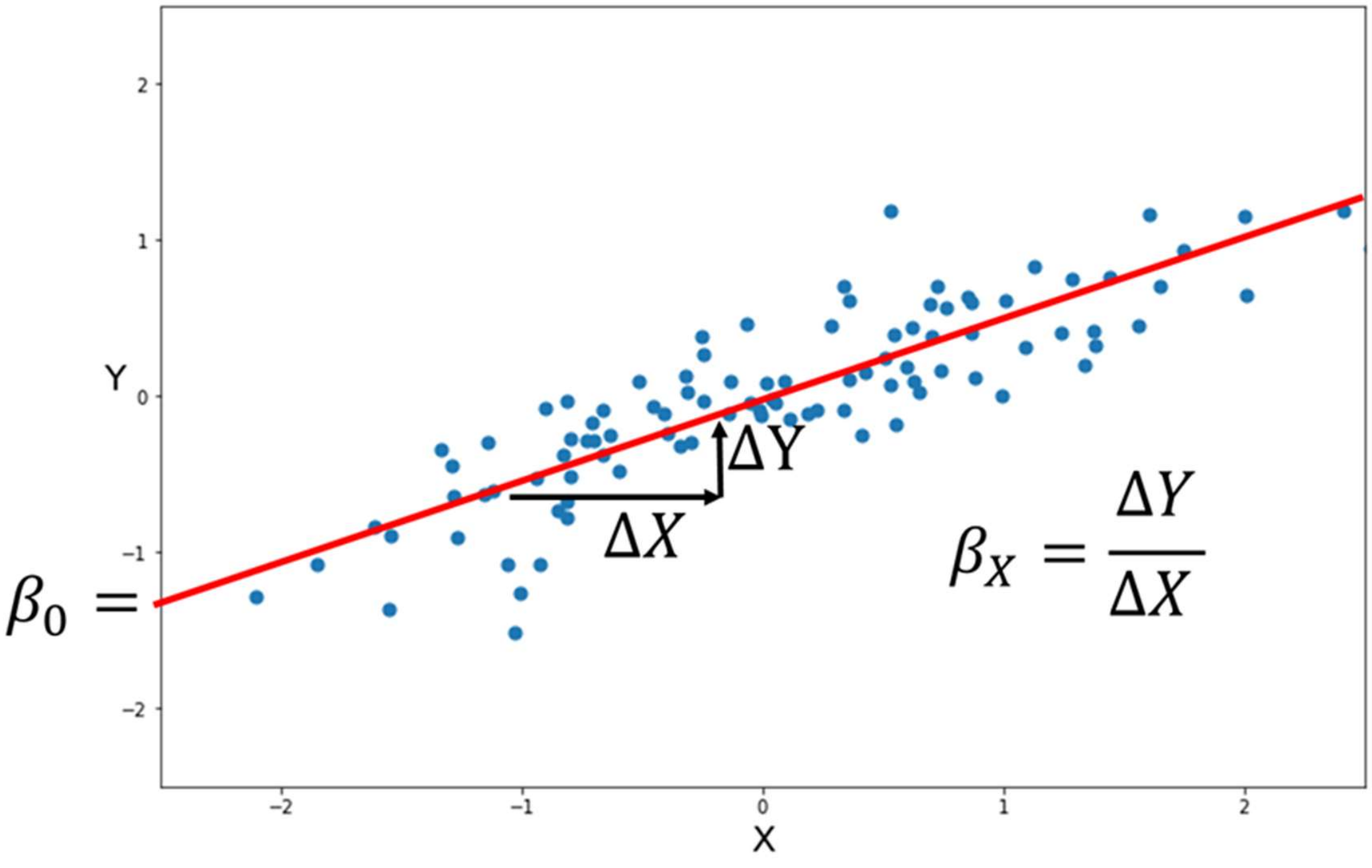

**Figure 1.** Scatter plot with a regression line fitted through the data. The regression coefficient corresponds to the slope of the regression line.

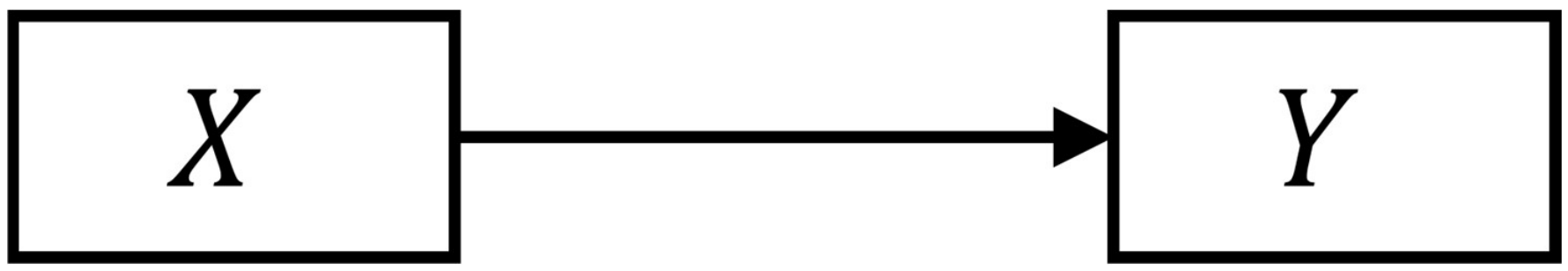

**Figure 2. Path diagram of the simple linear regression model.**

Before discussing some extensions of the regression model, please be aware that throughout this paper the error terms (ε) for the endogenous variables (i.e., variables which have an arrow pointing towards them) are not depicted in the Figures. In case of multiple predictors, a path model should specify correlations between the exogenous variables (i.e., those which do not have an arrow pointing towards them). For simplicity, these correlations will not be depicted in the Figures in this paper. Finally, the regression equations in this manuscript all include an intercept term for completeness, even when it can be assumed to be zero (for example after all variables are mean centred or standardized).

*2.2 Mediation analysis*

Mediation occurs when a variable $X$ indirectly influences another variable $Y$. The effect of the predictor $X$ is then said to be mediated by a third variable $M$, which is also a predictor of $Y$ (Shrout & Bolger, 2002). Figure 2 depicts a basic mediation model in which X affects variable M which in turn affects Y. The paths from $X$ to $M$ and from $M$ to $Y$ together represent the mediation. The strength of this indirect effect is calculated by multiplying the standardized regression coefficients belonging to these paths. (Baron & Kenny, 1986; Iacobucci et al., 2007; Keith, 2019). The path from $X$ to $Y$ indicates that there may be both a direct and an indirect effect of X.

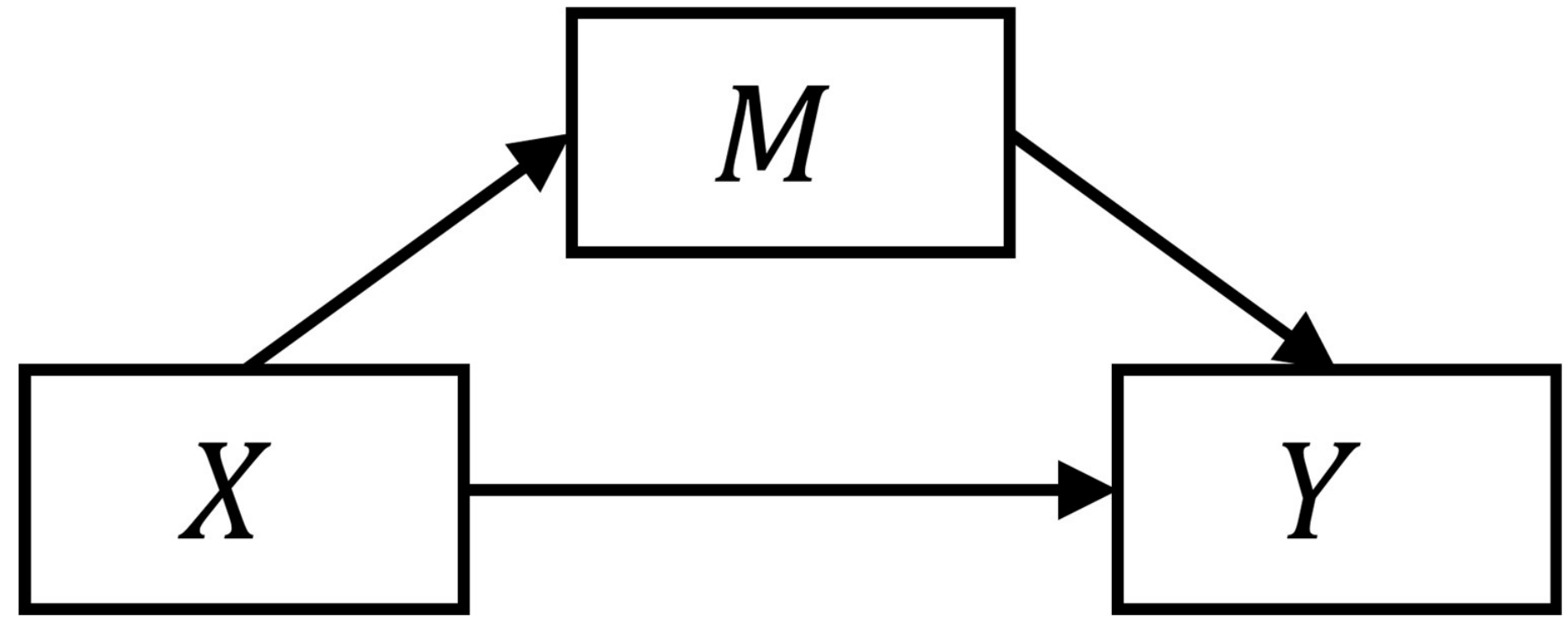

**Figure 2. Graphical representation of a basic (impure) mediation model.**

### 2.2.1 Assessment of mediation

The most common method to assess mediation, by Baron and Kenny (1986), is to fit three ordinary least squares regression models:

$$Y = \beta_{10} + \beta_{1X}X + \varepsilon_1 \quad (1)$$

$$M = \beta_{20}+\beta_{2X}X + \varepsilon_2 \quad (2)$$

$$Y = \beta_{30} + \beta_{3X}X + \beta_{3M}M + \varepsilon_3 \quad (3)$$

Mediation is determined when four conditions are met (MacKinnon et al., 2007; Muller et al., 2005):

1) The effect of X on Y in Equation 1 is significantly different from zero (i.e. $\beta_{1X} \neq 0$)
2) X has a significant effect on the mediator M (i.e. $\beta_{2X} \neq 0$, in Equation 2).
3) M has a significant effect on Y when controlling for X (i.e. $\beta_{3M} \neq 0$).
4) The direct effect of X on Y in Equation 3 must be smaller than the effect of X on Y in Equation 1 (i.e. $\beta_{3X} < \beta_{1X}$).

Edwards and Lambert (2007) added that when all conditions are met, but the estimate of $\beta_{3X}$ (i.e., $b_{3X}$) remains significantly different from zero, one concludes partial, or impure, mediation. When $b_{3X}$ has become non-significant, one concludes complete, or pure, mediation.

In complex models, with multiple mediators, Shrout and Bolger (2002) have shown that the first condition is not a necessary one and argued that it should be removed from the step-wise approach, as to protect the researcher from making a Type-II-error and losing power by discarding the mediation model as a whole if no direct effect is found.

*2.3 Moderation analysis*

In moderation, or interaction, the strength of the relationship between two variables depends on the value of a third variable (Morgan-Lopez & Mackinnon, 2006). In the conceptual representation of a moderation model in Figure 3, the path from $X$ to $Y$ is the main effect of interest. The arrow pointing towards the direct effect means that the effect of $X$ on $Y$ depends on (the value of) the moderator variable $Z$.

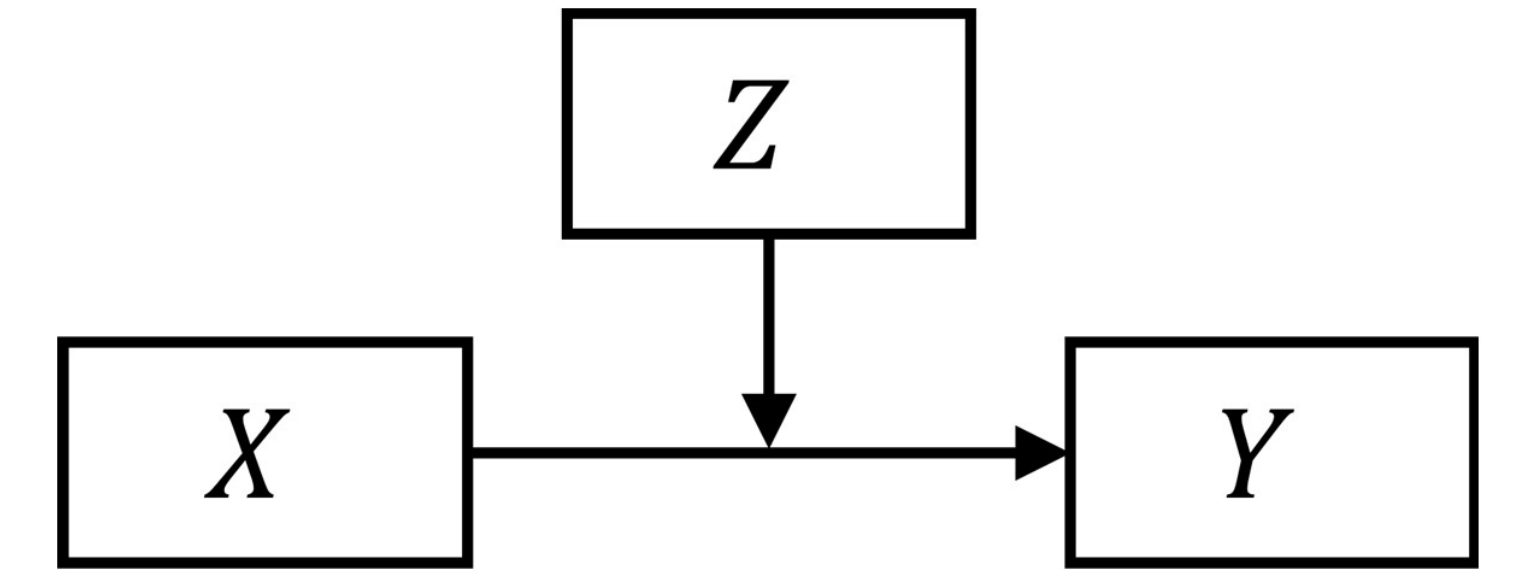

**Figure 3. Conceptual representation of a basic moderation model.**

2.4.1 Assessment of moderation

The general approach in moderation analysis to what is commonly called linear-by-linear interaction (Cohen et al., 2014) is to estimate a regression model in which the dependent variable Y is regressed on three predictors: X, the moderating variable Z and the product of Z and X (ZX).

$$Y = \beta_{40} + \beta_{4X}X + \beta_{4Z}Z + \beta_{4(ZX)}ZX + \varepsilon_4 \quad (4)$$

or

$$Y = [\beta_{40} + \beta_{4Z}Z] + [\beta_{4X} + \beta_{4(ZX)}Z]X + \varepsilon_4 \quad (4a)$$

By rewriting Equation 4 as a function of $X$ only, Equation 4a shows how the effect of X on Y is dependent on the value of $Z$. Luckily, assessment of moderation simply involves testing whether the regression parameter $\beta_{4(ZX)} \neq 0$.

An important note here is that the conceptual representation of the moderation model in Figure 3 does not correspond directly to the regression model of Equation 4. While some software implementations allow users to draw moderation models like the one in Figure 4, the path diagram related to Equation 4 is given in Figure 5.

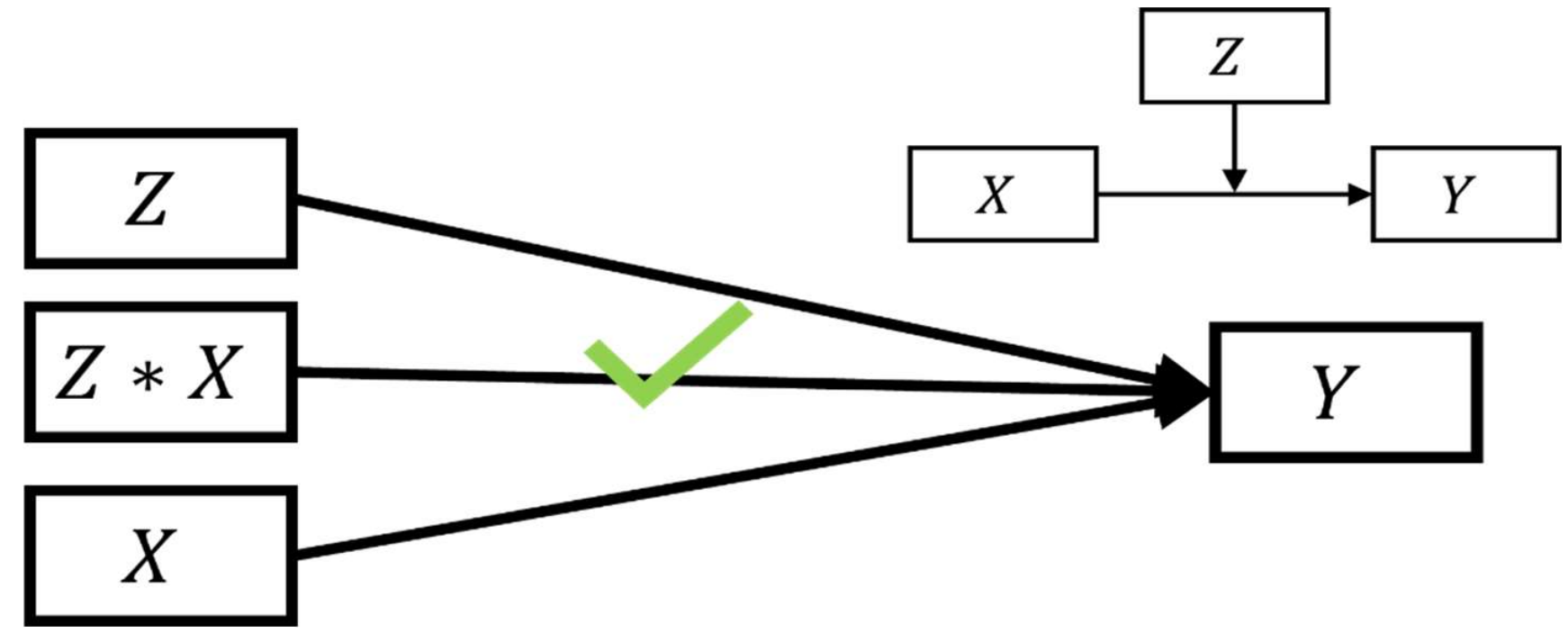

**Figure 4. Path model of a moderation model. If the regression coefficient for the arrow with the tick mark is significantly different from zero, moderation is assessed. In the top right is the conceptual representation.**

When doing moderation analysis, the 'main effects' now have a specific meaning. If a moderating effect is present, the parameters $b_{4X}$ and $b_{4Z}$ should only be interpreted as restricted conditional effects (Hayes & Matthes, 2009). Specifically, $b_{4X}$ is the difference expected in Y between two observations of which the first has a one-unit higher score on X, and both have scores zero on Z. When data are mean centred the estimate $b_{4X}$ is the expected difference in Y between two cases with a one-unit difference in X and mean scores on Z. In many cases therefore, the specific value of the main effect may not be that informative.

Apart from the added value of interpretation, mean centring can also decrease the correlation of lower order terms with their product-terms, thus decreasing non-essential multicollinearity (Cohen et al., 2014). Additionally, mean centring does not influence the parameter estimates for product terms (Croon, 2011). Analyses presented in this report therefore use mean-centred data.

*2.4 Integrations of mediation and moderation*

Current integrated models of mediation and moderation in the literature begin with a mediation model, where effects from and/or to the mediator are moderated by a fourth variable (Baron & Kenny, 1986; Edwards & Lambert, 2007; Morgan-Lopez & Mackinnon, 2006). In moderated mediation models, using the approach of Edwards and Lambert (2007), the model shown in Figure 6 represents a 'first and second-stage' moderated mediation, since both the effect to and from the mediator depend on the value of $Z$.

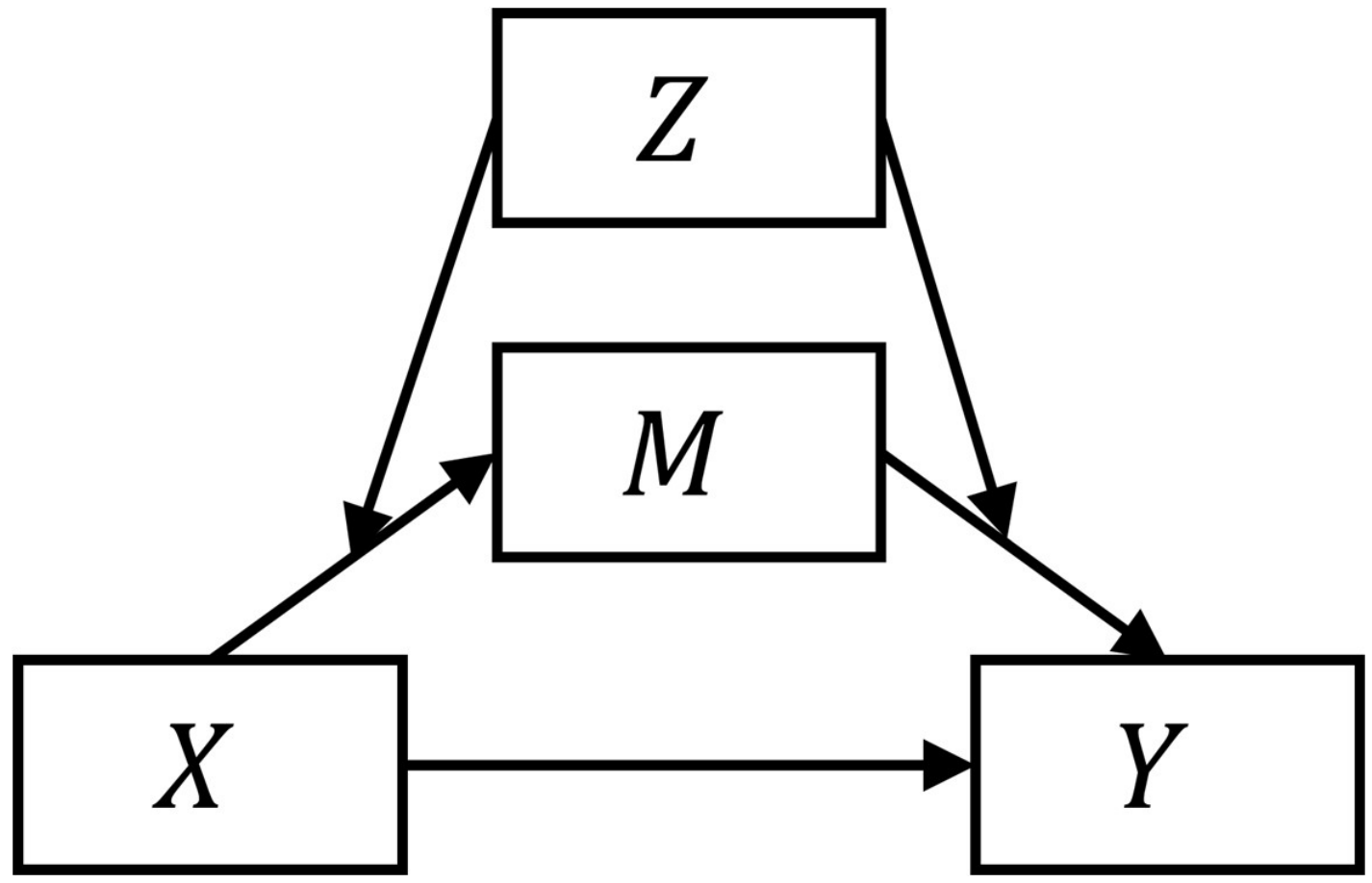

**Figure 5. First and second stage moderated mediation, as coined by Edwards and Lambert (2007).**

One case of the moderated mediation model has, rather confusingly, been called 'mediated moderation' (Morgan-Lopez & Mackinnon, 2006). In contrast to what the name implies (an indirect moderation) this model is analytically the same as a first stage moderated mediation model (see Figure 7) and thus contributes to the confusion about how to differentiate mediated moderation and moderated mediation (Edwards & Lambert, 2007).

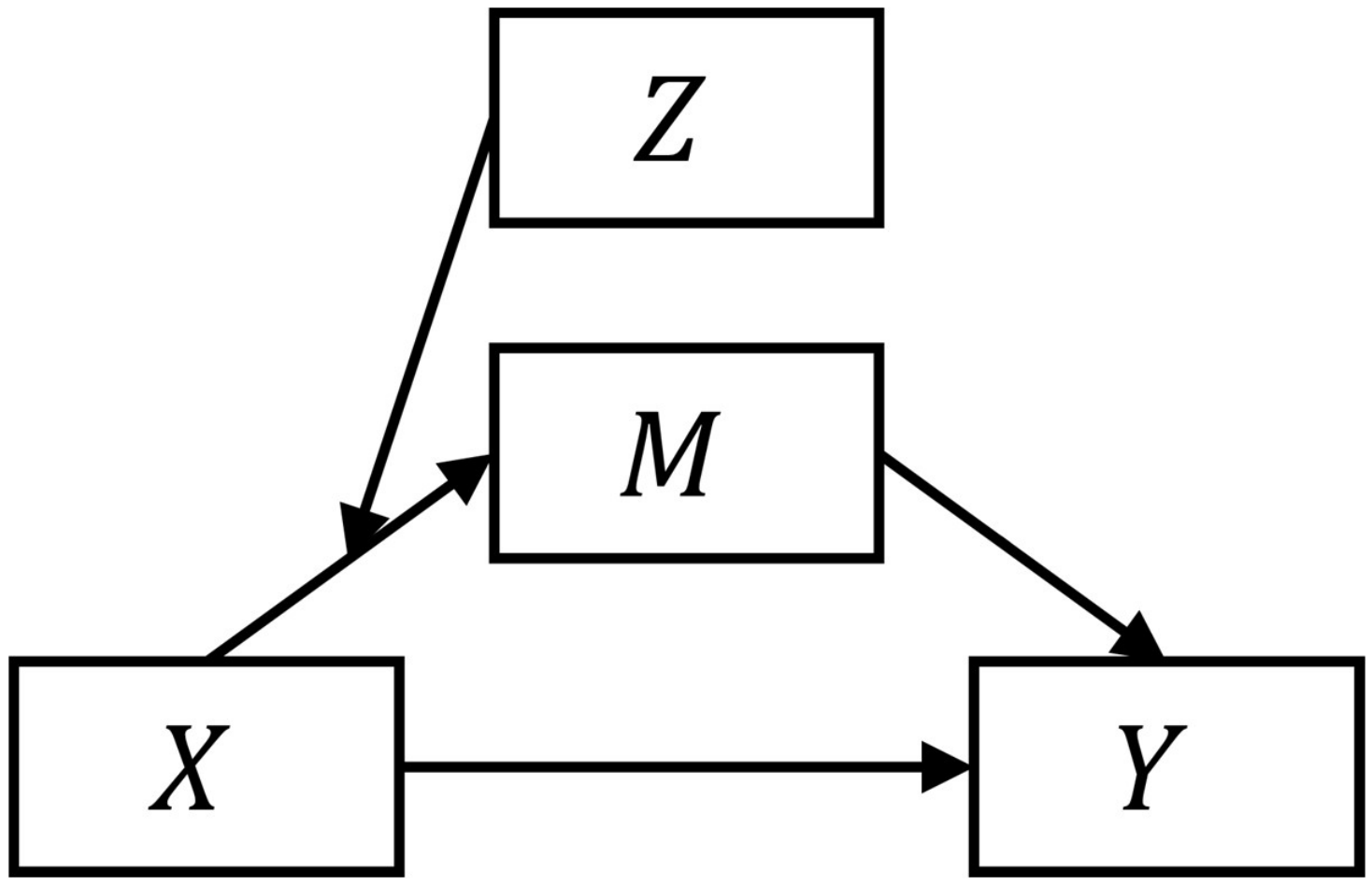

**Figure 6. The mediated moderation (first-stage moderated mediation) model of Baron and Kenny.**

*2.4.1 Assessment of the first stage 'mediated moderation' model*

The 'mediated moderation' model is assessed by checking whether a basic moderation model (Equation 4) can be also be described by a first stage moderated mediation model. Three regression models are used:

$$Y = \beta_{40} + \beta_{4X}X + \beta_{4Z}Z + \beta_{4(ZX)}ZX + \varepsilon_4 \quad (4)$$

$$M= \beta_{50} + \beta_{5X}X + \beta_{5Z}Z + \beta_{5(ZX)}ZX + \varepsilon_5 \quad (5)$$

$$Y= \beta_{60} + \beta_{6X}X + \beta_{6M}M + \beta_{6Z}Z + \beta_{6(ZX)}ZX+ \varepsilon_6 \quad (6)$$

Which correspond to the path model depicted in Figure 8.

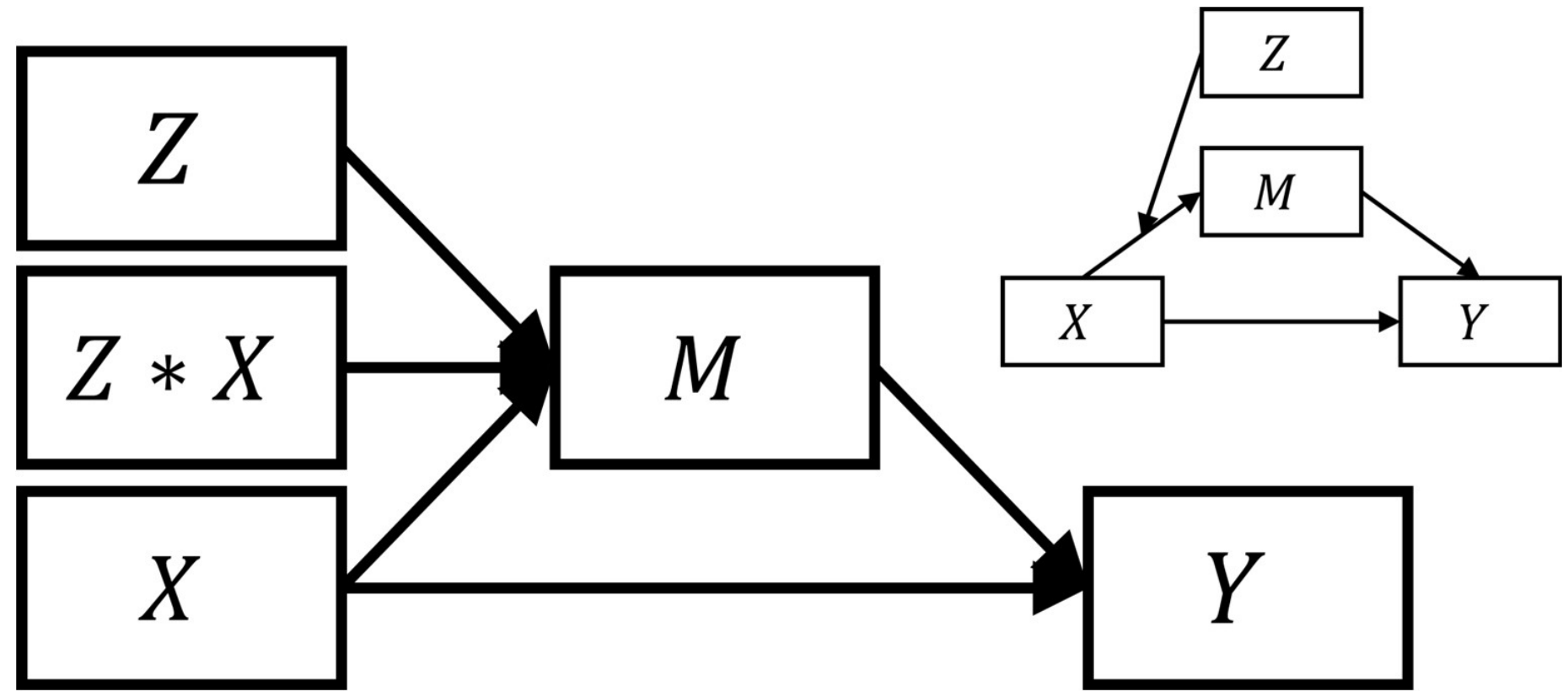

**Figure 7.Path model of the first stage mediated moderation model. In the top right the conceptual representation.**

The conditions for first-stage moderated mediation (a.k.a., 'Mediated moderation') are (Morgan-Lopez & Mackinnon, 2006):

1) $b_{5(ZX)}$ is significant
2) $b_{6(ZX)}$ is smaller in absolute value than $b_{4(ZX)}$

A more stringent method would be to test for the difference in the moderating effect between Equation 4 and 6 (i.e. is $\beta_{6(ZX)} < \beta_{4(ZX)}$ after inclusion of $M$).

*2.5 The indirect moderation model.*

The missing link in the regression framework is a model in which moderation is indirect (rather than an indirect effect being moderated). To avoid further confusion, the model will be referred to as 'indirect moderation'. Figure 8 shows how, in the indirect moderation model, a moderating effect of the variable $Z$ is indirect, via the actual true moderator variable $W$.

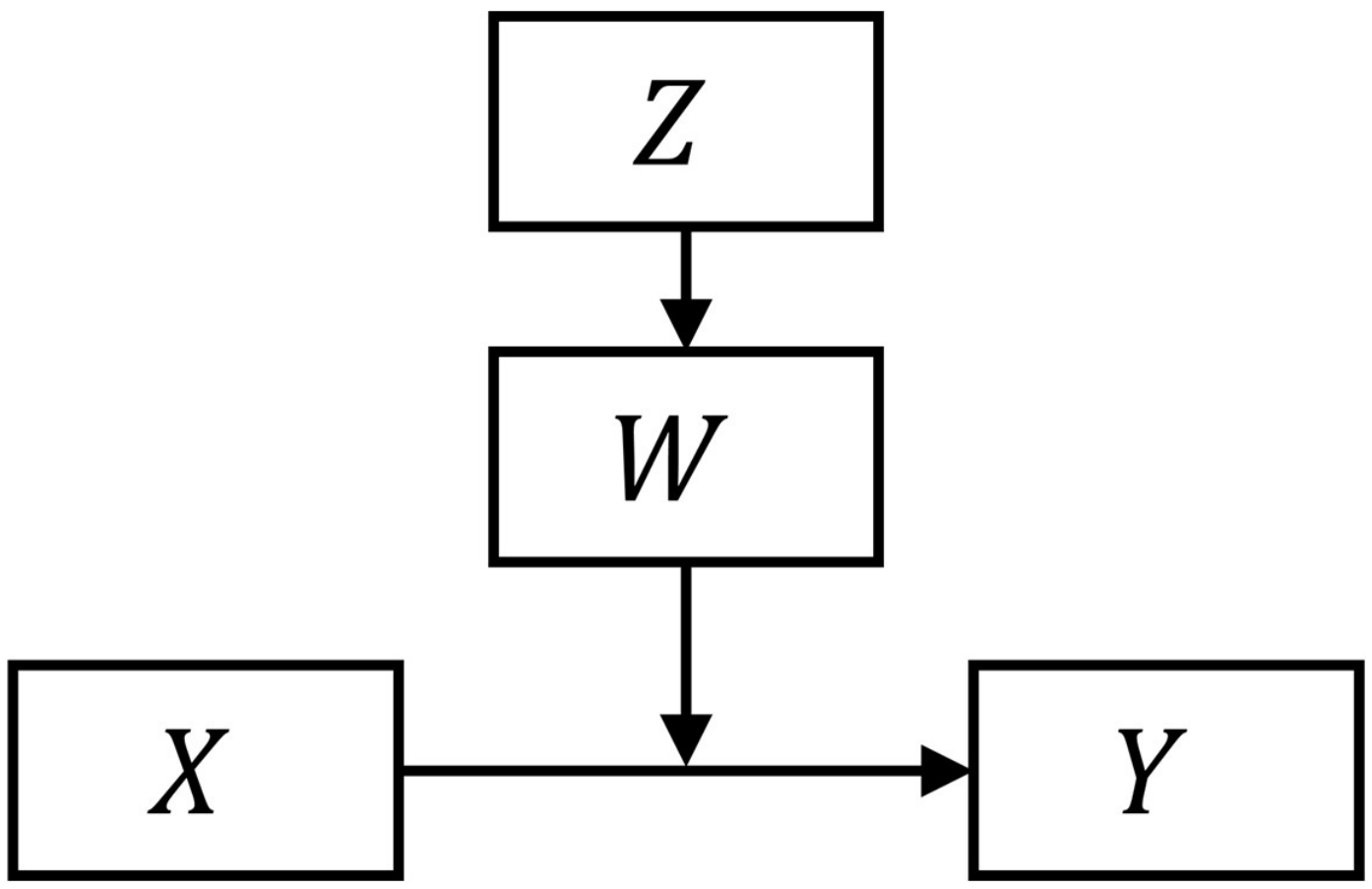

**Figure 8 Conceptual representation of indirect moderation. Due to the relation between Z and W, Z may be classified as a moderator if W is not included in the model.**

*2.5.1 Assessment of indirect moderation*

Logically incorporating mediation analysis in a moderation model means that assessment of indirect moderation requires 3 regression models. The letter W is used instead of M to indicate the functional difference of the variable in other models.

$$Y = \beta_{40} + \beta_{4X}X + \beta_{4Z}Z + \beta_{4(ZX)}ZX + \varepsilon_4 \quad (4)$$

$$Y = \beta_{70} + \beta_{7X}X + \beta_{7Z}Z + \beta_{7(ZX)}ZX + \beta_{7W}W + \beta_{7(WX)}WX + \varepsilon_7 \quad (7)$$

or

$$Y = \beta_{70} + [\beta_{7Z}Z + \beta_{7W}W] + [\beta_{7X} + \beta_{7(ZX)}Z+ \beta_{7(WX)}W]X + \varepsilon_7 \quad (7a)$$

$$W = \beta_{80} + \beta_{8Z}Z+ \beta_{8X}X+ \varepsilon_8 \quad (8)$$

The path model of the indirect moderation model, corresponding to the Equations 4, 7, and 8 is shown in Figure 9.

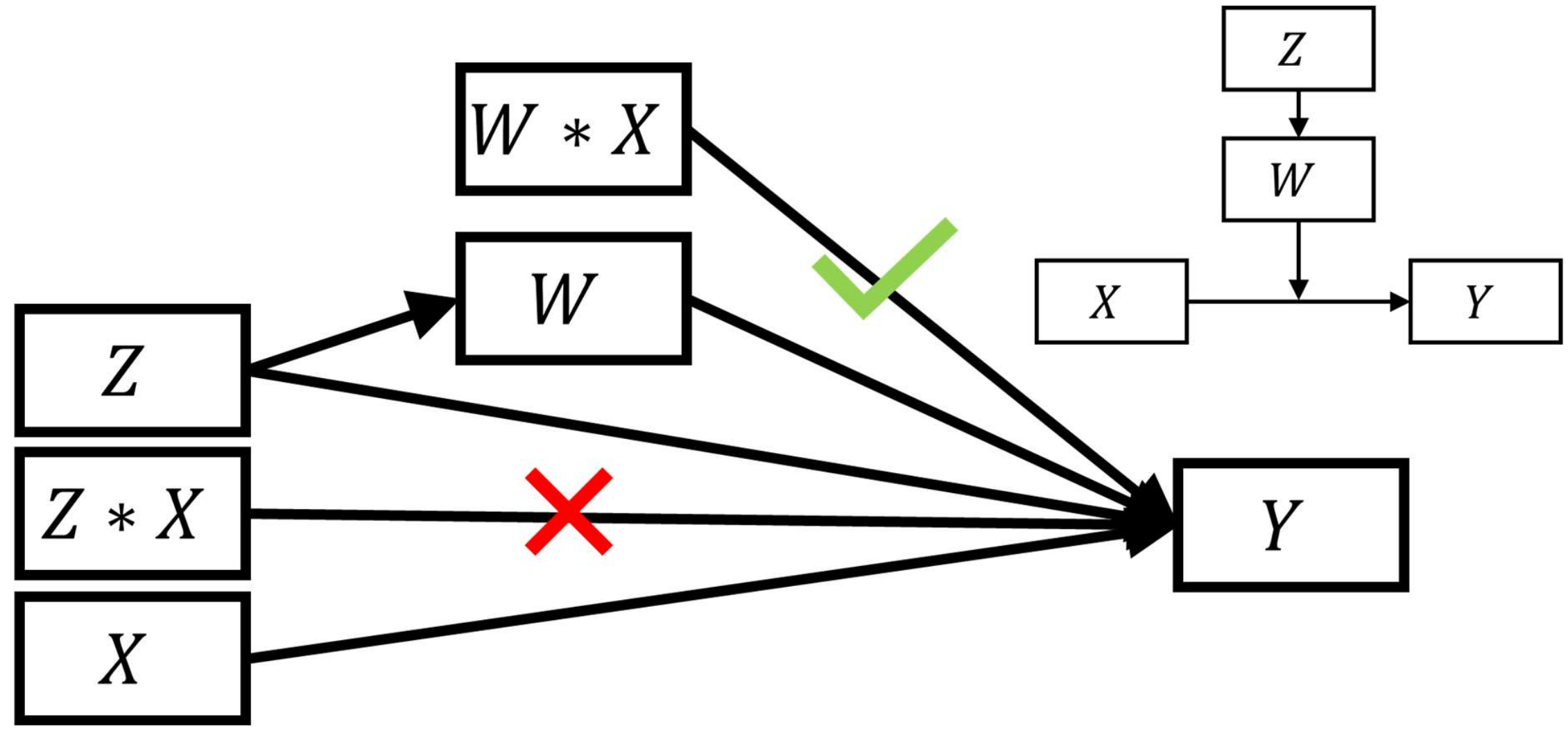

**Figure 9. Path model related to the indirect moderation model. On the right is the conceptual representation.**

Following the same logic of testing mediation and moderation, indirect moderation is assessed if:

1) there is a moderating effect of Z without considering W (i.e. $\beta_{4(ZX)} \neq 0$)
2) the moderating effect of Z is not present when W is included as a moderator (i.e. $\beta_{7(ZX)} = 0$),
3) Z has an effect on W (i.e. $\beta_{8Z} \neq 0$)
4) W moderates the effect of X on Y (i.e. $\beta_{7(WX)} \neq 0$)

Figure 10 provides a more conceptual decision tree.

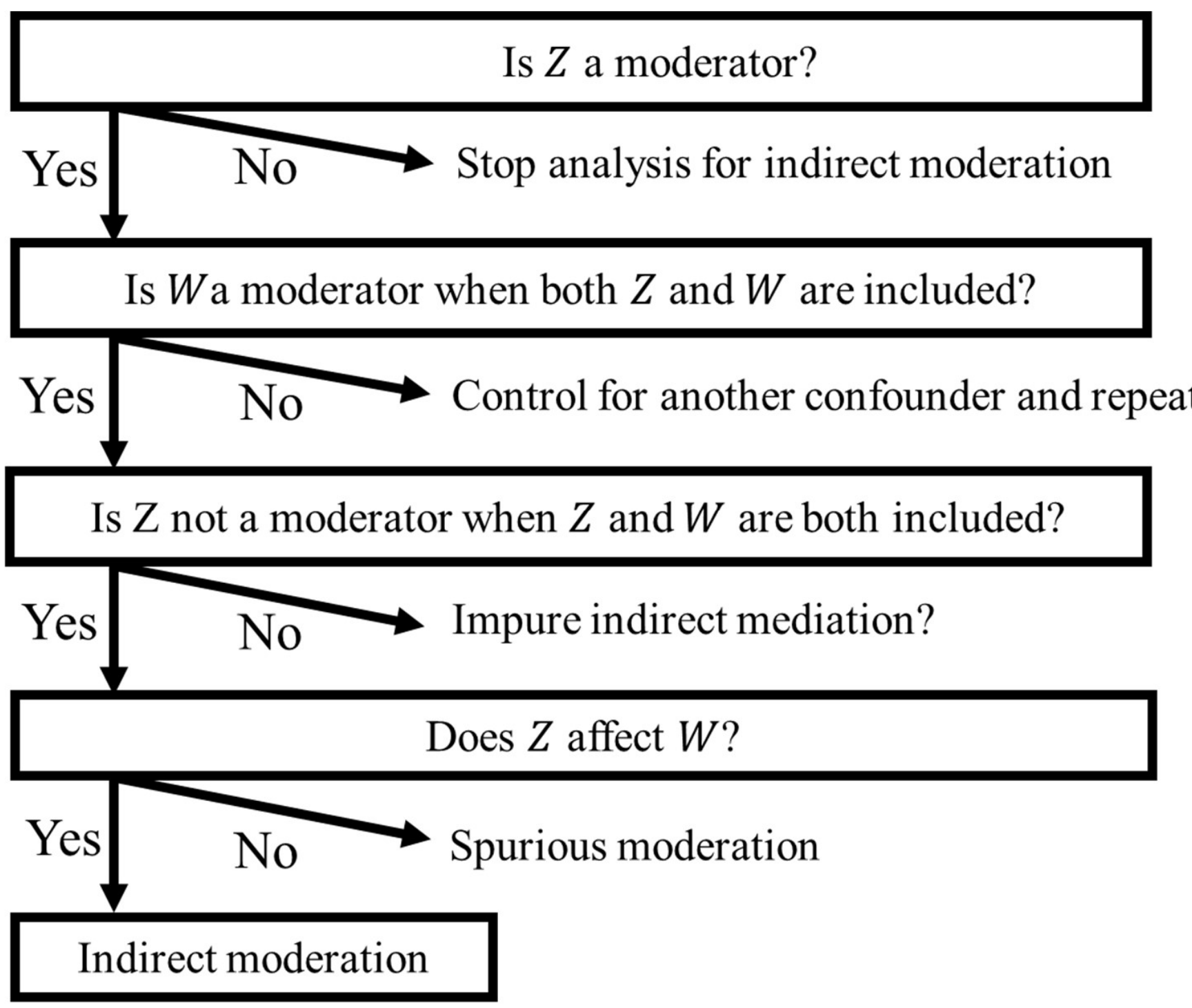

**Figure 9. Decision tree for assessing indirect moderation using stepwise regression analysis.**

The method described above is used to detect pure indirect moderation. Impure indirect moderation would have the looser restriction $|\beta_{5(ZX)}| < |\beta_{4(ZX)}|$. One idea is to adapt the method by Clogg *et al.*, (1995) to test for a significant change in the same parameter in two nested models. In such cases, an arrow from Z to the effect of X on Y could also be drawn in the conceptual representation.

Since Equation 4 is nested in Equation 7 (i.e. Equation 4 is a special case of Equation 7 by assuming that $\beta_{7W}$=0 and $\beta_{7(WX)}$ =0) one might want to use a test for model comparison (e.g. an $R^2$-change test). However, Equation 7 does not have to explain significantly more variance than Equation 4 to be informative, as the goal should be to explain underlying processes better substantively. If the effect of ZX ($b_{7(ZX)}$) becomes non-significant after the inclusion of the W terms, maybe not more variance is explained, but it is explained in a better way.

There are several situations, based on the indirect moderation model, where W can be a mediator of the moderating effect of Z, all of which are based on the three characteristics that Z has an effect on W, Z is a moderator of the effect of X on Y when W is not taken into account, and the moderating effect of Z is weaker when W is taken into account. The discussion here will, however, be restricted to linear-by-linear moderation of continuous variables and the situation in which W is a pure mediator of the moderating effect of Z (i.e. Z does not have an additional moderating effect in the population).

2.5.2 Special cases.

When using this decision-tree one will obviously decide ‘no’ in some cases. However, all is not lost when some criteria for moderated mediation are not found. When in Step 1 no moderating effect of Z is found, one can stop the analysis for mediated moderation. When in the next step W is not a significant moderator, one could use another confounding moderator and repeat Step 2 with another. Note that choosing these confounders should be based on theoretical rather than statistical arguments.

When another moderator is found and in Step 3 of the decision-tree and Z still remains significant, one may conclude a multiple moderator model, which may again undergo the same prescribed steps for assessing mediated moderation, and assess impure (or partial) indirect moderation. Analysis of such models may be investigated further but surpass the scope of the current foundational paper.

If, however, Z does fail to be a moderator when W is included as a moderator but there is no effect of Z on W, one will conclude that there is a spurious (non-authentic) moderating effect of Z. Not all these possibilities have been explicitly recorded. However, the reader might be able to infer the quantitative data for some of these alternatives from the results described below and by running the simulation with different criteria (syntax for the study below is provided as supplemental material).

## 3. **Simulation experiment**

### *3.1 Simulation setup*

To study the behaviour of the decision-tree, data was generated and analysed using the open-source programming software 'R' (Team R Development Core, 2018), with the data generating process being:

$$Y = \beta_X X + \beta_Z Z + \beta_W W + \beta_{WX} WX + \varepsilon_Y, \text{ with } \varepsilon_Y \sim \text{Normal}(0, 1) \quad (9)$$

The relationships between the three continuous variables were manipulated to determine the effects on the regression parameter estimates of interest in the decision-tree. For datasets with sample sizes of $N \in \{100, 150\}$, the effects of parameter values for$\beta_{WX} \in \{-.4, -.2, 0, .2, .4\}$ were studied. All other regression coefficients were set to .3. The effect of the correlation between Z and W was studies by varying the correlation $\rho_{ZW} \in \{0, .3, .6\}$. The correlations between X and W and between X and Z were set constant at .4. The full R code to reproduce this study is provided as supplemental material.

For these 2*5*3=30 conditions the results will be reported in two ways. Firstly, the results will be given in marginal proportions of 'yes' answers to each step in the decision tree. This will allow for detailed evaluation of Type I error rates and power to detect certain effects. Secondly, the results are provided in conjunctional form. That is, it is evaluated how often the steps of the decision-tree were successively answered with 'yes'. After four successive 'yes' answers, indirect moderation is assessed.

*3.2 Expectations*

It is important to, a priori, determine relationships between the parameters, based on the relationships between the variables. Since the method starts with inclusion of the wrong moderator, it is helpful to investigate how the model parameters are related. Specifically, how other factors influence the parameter $\beta_{4ZX}$ in Equation 4. Using Mathematica 8.0 (Wolfram Research Inc., 2010) it was derived that in a population model for three independent variables:

$$\beta_{4ZX} = \beta_{WX} * \frac{\sigma_W}{\sigma_Z} * \frac{\rho_{ZW} + \rho_{WX}\rho_{ZX}}{1 + \rho_{ZX}^2} \quad (10)$$

where $\beta_{WX}$ is the moderating effect of W in the true model, σ represents the standard deviation of the variable noted in the subscript and ρ refers to a specific element from the correlation matrix R

| | Z | W | X |
|---|---|---|---|
| Z | 1 | $\rho_{ZW}$ | $\rho_{ZX}$ |
| W | $\rho_{ZW}$ | 1 | $\rho_{WX}$ |
| X | $\rho_{ZX}$ | $\rho_{WX}$ | 1 |

For completeness: the mathematical result only holds when applied to population data and under the assumption that W is the true moderator, while Z is treated as moderator instead and all variables are normally distributed.

Equation 10 implies that one can expect to find significant moderator effects of Z more often than can be expected from the chosen level of significance. For this study, the convenient level of significance, or expected Type I error rate, of .05 (Fisher, 1939) was used for all parameters estimates in our simulations. Note that this Type I error rate will be inflated (i.e., larger than expected) when the:

- moderating effect of W increases,
- variance of W increases,
- variance of Z decreases,
- correlation between Z and W increases,
- correlation between Z and X decreases,
- product of the correlations between (W and Z) and (Z and X) increases

It is a straightforward prediction that the Type I error rate for the regression parameter for the moderating effect of Z will be inflated (i.e. a significant moderating effect of Z will be found more than 5% of the time) in the study described above as a function of the moderating effect of W and the correlation $\rho_{ZW}$. This prediction is justified since all other variables from Equation 10 are kept constant throughout our study.

## 4. **Results and discussion**

Table 1 gives the primary distributions of the steps. The elements give the proportion of the 10,000 simulations which gave a 'yes' answer to each question irrespective of the other results. For example, the rows where $\beta_{WX} = 0$ show that Type I error rates are nominal (i.e., close to the conveniently chosen significance level $\alpha = .05$) for all steps in the decision tree (note that the negation in Step 3 makes that .95 is the positive outcome). Indirect moderation was nearly never wrongfully assessed.

**Table 1.** Primary distributions of 'yes' answers for each condition, irrespective of the results of other conditions.

| N | βWX | ρZW | Step 1 $\beta_{4ZX} \neq 0$ | Step 2 $\beta_{7WX} = 0$ | Step 3 $\beta_{7ZX} \neq 0$ | Step 4 $\beta_{8Z} \neq 0$ | Indirect Moderation |
|---|---|---|---|---|---|---|---|
| 100 | -0,4 | 0 | 0.118 | 0.961 | 0.948 | 0.052 | 0.010 |
| | | 0,3 | 0.327 | 0.934 | 0.952 | 0.870 | 0.271 |
| | | 0,6 | 0.661 | 0.834 | 0.948 | 1.000 | 0.539 |
| | -0,2 | 0 | 0.071 | 0.500 | 0.949 | 0.052 | 0.002 |
| | | 0,3 | 0.130 | 0.454 | 0.951 | 0.871 | 0.049 |
| | | 0,6 | 0.262 | 0.334 | 0.950 | 1.000 | 0.088 |
| | 0 | 0 | 0.053 | 0.047 | 0.946 | 0.053 | 0.000 |
| | | 0,3 | 0.048 | 0.053 | 0.953 | 0.865 | 0.001 |
| | | 0,6 | 0.049 | 0.050 | 0.951 | 1.000 | 0.001 |
| | 0,2 | 0 | 0.065 | 0.501 | 0.953 | 0.050 | 0.003 |
| | | 0,3 | 0.133 | 0.454 | 0.949 | 0.862 | 0.050 |
| | | 0,6 | 0.254 | 0.326 | 0.948 | 1.000 | 0.083 |
| | 0,4 | 0 | 0.118 | 0.955 | 0.951 | 0.053 | 0.010 |
| | | 0,3 | 0.317 | 0.937 | 0.952 | 0.862 | 0.264 |
| | | 0,6 | 0.668 | 0.828 | 0.948 | 1.000 | 0.542 |
| 250 | -0,4 | 0 | 0.173 | 1.000 | 0.948 | 0.051 | 0.012 |
| | | 0,3 | 0.635 | 1.000 | 0.950 | 0.999 | 0.609 |
| | | 0,6 | 0.964 | 0.997 | 0.947 | 1.000 | 0.919 |
| | -0,2 | 0 | 0.084 | 0.895 | 0.954 | 0.048 | 0.004 |
| | | 0,3 | 0.252 | 0.843 | 0.950 | 0.998 | 0.197 |
| | | 0,6 | 0.562 | 0.697 | 0.950 | 1.000 | 0.387 |
| | 0 | 0 | 0.051 | 0.050 | 0.951 | 0.050 | 0.000 |
| | | 0,3 | 0.053 | 0.049 | 0.948 | 0.999 | 0.001 |
| | | 0,6 | 0.051 | 0.052 | 0.951 | 1.000 | 0.001 |
| | 0,2 | 0 | 0.084 | 0.891 | 0.951 | 0.050 | 0.004 |
| | | 0,3 | 0.251 | 0.848 | 0.951 | 0.998 | 0.202 |
| | | 0,6 | 0.553 | 0.699 | 0.951 | 1.000 | 0.384 |
| | 0,4 | 0 | 0.169 | 1.000 | 0.949 | 0.049 | 0.011 |
| | | 0,3 | 0.641 | 1.000 | 0.951 | 0.999 | 0.611 |
| | | 0,6 | 0.963 | 0.997 | 0.949 | 1.000 | 0.920 |

The results are in line with expectations that, depending on the correlation between the proposed and the true moderator, the probability of making a Type I error for the wrong moderator becomes greatly inflated. Importantly the Type I error rate is a function of strength of the moderating effect of W (even while that predictor is not included in the regression model).

The results also show a detrimental effect of including a correlated non-moderating variable as a moderator on the probability of finding a significant result for the true moderator. That is, the Type II error rate increases with the correlation between W and Z. Next to the mathematical proof, this behaviour can be seen in the second column of the primary results. The number of significant results diminishes when the correlation between W and Z increases, keeping all other things constant.

When sample size increases, the probability of finding a significant result for the moderating effect of Z also increases, thus making more Type I errors. The overall probabilities of finding a significant result for the moderating effect of W increases with sample size. As far as these results go, the effect of including both Z and W as moderators on the probability of finding a significant result for the moderating effect of Z does not depend on sample size. For the effect of Z on W, the same parameter behaviour was observed in both sample sizes, albeit that larger sample sizes overall have more significant results. Hence, not surprisingly, the probability of correctly assessing indirect moderation increases with the sample size.

The entire column belonging to Step 3 shows that the Type I error rate is nominal in all conditions. The powerful implication is that, regardless of whether a variable (in this case Z) is wrongly classified as a moderator at first, including the true moderator will control for the faulty result. Covariates should therefore also be included in regression models to control for spurious moderation, and not only for spurious main effects.

Instead of evaluating the Steps separately, Table 2 gives the conjunctive distributions for reference in proportion of 10,000 simulations. The values in the Table indicate how often the questions up to and including that Step were answered with ‘yes’. For example: the proportion .087 in the first row, column for Step 3, indicates that in 870 simulations Step 3 was answered with ‘yes’ when Step 1 and 2 were also answered with ‘yes’ (Table 1 shows that the total proportion of ‘yes’ answers for Step 3, was .948). The conjunctional distribution in the first column (Step 1) of Table 2 is the same as the marginal distribution in the first column (Step 1) of Table 1.

**Table 2.** Conjunctional distribution of each successive 'yes' answer. The proportions in the last two columns are the same.

| N | $\beta_{WX}$ | $\rho_{ZW}$ | Step 1<br>$\beta_{4ZX} \neq 0$ | Step 2<br>$\beta_{7WX} = 0$ | Step 3<br>$\beta_{7ZX} \neq 0$ | Step 4<br>$\beta_{8Z} \neq 0$ | Indirect Moderation |
|---|---|---|---|---|---|---|---|
| 100 | -0,4 | 0 | 0.118 | 0.113 | 0.087 | 0.010 | 0.010 |
| | | 0,3 | 0.327 | 0.306 | 0.285 | 0.271 | 0.271 |
| | | 0,6 | 0.661 | 0.551 | 0.539 | 0.539 | 0.539 |
| | -0,2 | 0 | 0.071 | 0.037 | 0.023 | 0.002 | 0.002 |
| | | 0,3 | 0.130 | 0.058 | 0.052 | 0.049 | 0.049 |
| | | 0,6 | 0.262 | 0.089 | 0.088 | 0.088 | 0.088 |
| | 0 | 0 | 0.053 | 0.002 | 0.001 | 0.000 | 0.000 |
| | | 0,3 | 0.048 | 0.001 | 0.001 | 0.001 | 0.001 |
| | | 0,6 | 0.049 | 0.002 | 0.001 | 0.001 | 0.001 |
| | 0,2 | 0 | 0.065 | 0.034 | 0.022 | 0.003 | 0.003 |
| | | 0,3 | 0.133 | 0.061 | 0.054 | 0.050 | 0.050 |
| | | 0,6 | 0.254 | 0.084 | 0.083 | 0.083 | 0.083 |
| | 0,4 | 0 | 0.118 | 0.113 | 0.088 | 0.010 | 0.010 |
| | | 0,3 | 0.317 | 0.299 | 0.281 | 0.264 | 0.264 |
| | | 0,6 | 0.668 | 0.554 | 0.542 | 0.542 | 0.542 |
| 250 | -0,4 | 0 | 0.173 | 0.173 | 0.144 | 0.012 | 0.012 |
| | | 0,3 | 0.635 | 0.634 | 0.609 | 0.609 | 0.609 |
| | | 0,6 | 0.964 | 0.962 | 0.919 | 0.919 | 0.919 |
| | -0,2 | 0 | 0.084 | 0.075 | 0.051 | 0.004 | 0.004 |
| | | 0,3 | 0.252 | 0.213 | 0.197 | 0.197 | 0.197 |
| | | 0,6 | 0.562 | 0.390 | 0.387 | 0.387 | 0.387 |
| | 0 | 0 | 0.051 | 0.002 | 0.001 | 0.000 | 0.000 |
| | | 0,3 | 0.053 | 0.002 | 0.001 | 0.001 | 0.001 |
| | | 0,6 | 0.051 | 0.003 | 0.001 | 0.001 | 0.001 |
| | 0,2 | 0 | 0.084 | 0.075 | 0.049 | 0.004 | 0.004 |
| | | 0,3 | 0.251 | 0.217 | 0.202 | 0.202 | 0.202 |
| | | 0,6 | 0.553 | 0.388 | 0.384 | 0.384 | 0.384 |
| | 0,4 | 0 | 0.169 | 0.169 | 0.142 | 0.011 | 0.011 |
| | | 0,3 | 0.641 | 0.641 | 0.612 | 0.611 | 0.611 |
| | | 0,6 | 0.963 | 0.960 | 0.920 | 0.920 | 0.920 |

## 5. ***Comparing model fit of indirect moderation and 'mediated moderation'***

Experience has shown that researchers tend to perceive the indirect moderation model as being the same as the Baron and Kenny (1986) model named 'mediated moderation', even after explaining the differences. To solve this, one approach is to show that each model has a different fit to data generated under the indirect moderation model. To first deduce the algebraic differences between the models, assume that M and W are the same variable but play a different role each model. Deriving the regression equation for predicting Y in the Baron and Kenny model results in the equation:

$$Y_{BK} = \beta_{Y0} + \beta_{YM}\beta_{M0} + (\beta_{YX} + \beta_{YM}\beta_{MX})X + (\beta_{YM}\beta_{MZ})Z + (\beta_{YM}\beta_{MZX})ZX + (\beta_{YM})\varepsilon_M + \varepsilon_Y \quad (11)$$

Similarly, the regression for predicting Y in the indirect moderation model can be written as

$$Y_{IndMo} = \beta_{Y0} + \beta_{YW}\beta_{W0} + (\beta_{YX} + \beta_{YWX}\beta_{W0})X + (\varepsilon_W\beta_{YWX})X + (\beta_{YZ} + \beta_{YW}\beta_{WZ})Z + (\beta_{YWX}\beta_{WZ})ZX + \varepsilon_Y \quad (12)$$

The models for predicting Y are not identical and one crucial difference is that the indirect moderation model includes a random effect of X. That is, the regression parameter of X depends on the random component $\varepsilon_W$ (Christensen, 1997)

To test whether the models are statistically equivalent or not, and thus whether or not they can answer different research questions both models were fitted using AMOS 18 (Arbuckle, 2009) to the covariance matrix of a data set generated under our model with the variable parameters set to n=250, $\beta_{WX} = -0.2$ and $\rho_{ZW} = 0.3$. If the models are equivalent, the same fit statistics would be observed for both models.

The results of these analyses (see Table 3) show that the indirect model fitted the data very acceptably, while the Baron and Kenny (1986) model did not. The probability that our model fitted the data perfectly was .617 ($\chi^2$ = 0.966 with df = 2). Also the descriptive fit indices provided very acceptable results (Tabri & Elliott, 2012). The Baron and Kenny model however did not fit the data well. The probability that the Baron and Kenny model fitted the data perfectly was .003 ($\chi^2$ = 11.662 with df = 2) and the descriptive fit indices indicated poor fit.

**Table 3. Fit statistics from the SEM analysis of 'mediated moderation' and the indirect moderation model. AGFI and TLI values greater than .95 are considered acceptable. TLI values greater than 1 are often set to 1 to ease interpretation. AGFI values dependent on sample size and can therefore only be used as a comparison between models for the same data**

| | Statistical indices | | | Descriptive | |
|---|---|---|---|---|---|
| | $\chi^2$ | df | p-value | AGFI | TLI |
| Indirect moderation | 0.966 | 2 | 0.617 | 0.986 | 1.029 |
| Baron and Kenny model | 11.662 | 2 | 0.003 | 0.842 | 0.730 |

The SEM analysis shows that the models were statistically non-equivalent and may be used to answer different research questions. As was argued in the beginning, the indirect moderation model is not just a special case of moderated mediation as is the Baron and Kenny model.

## 5 Conclusions

In this tutorial, the indirect moderation model was presented which can be used to analyse processes where a variable moderates an effect through another variable. A simulation study showed that the Type I error rate in moderation analysis can be high if a variable is included as a moderator if that variable is only related to the true moderator. The solution is simple: control for spurious moderators by including control variables. When one of the two proposed variables is not a moderator, this will be found in Step 3 of the decision tree. Also, the inclusion of two suspected moderator variables has been shown to be a robust way of determining which one is the true moderator. Future work may investigate the robustness of the methodology under different conditions (e.g., with multiple control moderators simultaneously).

There was a strong decrease in power to detect a true moderator W in the presence of a wrongly included moderator. Low to moderately correlated variables can have detrimental effects of the reliability of regression estimates for the moderators. This effect should be carefully considered in research applications as low to moderate correlations often exist between many variables in the social and behavioural sciences.

It is extremely important for any researcher using moderators in regression analysis to become aware of the pitfalls of including the wrong moderator in a model. Researchers can gain much in research validity when they would not only include covariates as main effects, but also include possible confounding moderators. More research is necessary to investigate the behaviour of parameters of moderators in multiple moderator models, but this presented methodology may be an important step to answering new research questions.

Appendix A.

Derivations for comparing the regression models. W and M are assumed to essentially be the same variable but play distinct roles in each model.

<u>Baron and Kenny (BK) model</u>

$M \quad = \beta_{M0} + \beta_{MZ}Z + \beta_{MX} X + \beta_{MZX} ZX + \varepsilon_{M}$

$Y_{BK} \quad = \beta_{Y0} + \beta_{YX} X + \beta_{YM} M + \varepsilon_{Y}$

SUBSTITUTING FOR M GIVES

$Y_{BK} \quad = \beta_{Y0} + \beta_{YX} X + \beta_{YM}[\beta_{M0} + \beta_{MZ}Z + \beta_{MX} X + \beta_{MZX} ZX + \varepsilon_{M}] + \varepsilon_{Y}$

$= \beta_{Y0} + \beta_{YX} X + \beta_{YM}\beta_{M0} + (\beta_{YM}\beta_{MZ})Z + (\beta_{YM}\beta_{MX})X + (\beta_{YM}\beta_{MZX}) ZX + (\beta_{YM}) \varepsilon_{M} + \varepsilon_{Y}$

REWRITING THIS EXPRESSION GIVES

$Y_{BK} \quad = \beta_{Y0} + \beta_{YM}\beta_{M0} + (\beta_{YX} + \beta_{YM}\beta_{MX}) X + (\beta_{YM}\beta_{MZ})Z + (\beta_{YM}\beta_{MZX}) ZX + (\beta_{YM}) \varepsilon_{M} + \varepsilon_{Y}$

<u>Indirect Moderation model:</u>

$W \quad = \beta_{W0} + \beta_{WZ} Z + \varepsilon_{W}$

$Y_{IndMo} = \beta_{Y0} + \beta_{YX} X + \beta_{YZ} Z + \beta_{YW} W + \beta_{YWX} WX + \varepsilon_{Y}$

SUBSTITUTING FOR W GIVES

$Y_{IndMo} = \beta_{Y0} + \beta_{YX} X + \beta_{YZ} Z + \beta_{YW}[\beta_{W0} + \beta_{WZ} Z + \varepsilon_{W}] + \beta_{YWx}X[\beta_{W0} + \beta_{WZ} Z + \varepsilon_{W}] + \varepsilon_{Y}$

$= \beta_{Y0} + \beta_{YX} X + \beta_{YZ} Z + \beta_{YW}\beta_{W0} + (\beta_{YW}\beta_{WZ}) Z + (\beta_{YW}) e_{W} + (\beta_{YWX} \beta_{W0}) X + (\beta_{YWx}\beta_{WZ})XZ + \varepsilon_{W}(\beta_{YWx}X) + \varepsilon_{Y}$

REWRITING THE EXPRESSION GIVES

$Y_{IndMo} = \beta_{Y0} + \beta_{YW}\beta_{W0} + (\beta_{YX} + \beta_{YWX} \beta_{W0}) X + (\varepsilon_{W}\beta_{YWX})X + (\beta_{YZ} + \beta_{YW}\beta_{WZ})Z + (\beta_{YWx}\beta_{WZ})ZX + +(\beta_{YW}) e_{W} + \varepsilon_{Y}$

<u>COMPARISON OF THE TWO REGRESSIONS ON Y</u>

$Y_{BK} \quad = \beta_{Y0} + \beta_{YM}\beta_{M0} + (\beta_{YX} + \beta_{YM}\beta_{MX}) X + (\beta_{YM}\beta_{MZ})Z + (\beta_{YM}\beta_{MZX}) ZX + (\beta_{YM}) \varepsilon_{M} + \varepsilon_{Y}$

$Y_{MeMo} = \beta_{Y0} + \beta_{YW}\beta_{W0} + (\beta_{YX} + \beta_{YWX} \beta_{W0} + \varepsilon_{W}\beta_{YWX}) X + (\beta_{YZ} + \beta_{YW}\beta_{WZ})Z + (\beta_{YWx}\beta_{WZ})ZX + (\beta_{YW}) e_{W} + \varepsilon_{Y}$